\documentclass[12pt,preprint]{emulateapj}

\slugcomment{Version date: \today}

\shorttitle{VLA Survey of the CDFS. IV. Source Population}
\shortauthors{Padovani et al.}

\begin{document}
\def\proptoapprox{\mathrel{\hbox{\rlap{\hbox{\lower4pt\hbox{$\sim$}}}\hbox{$\propto$}}}}
\newdimen\digitwidth
\setbox0=\hbox{2}
\digitwidth=\wd0
\catcode `#=\active
\def#{\kern\digitwidth}

\title{The VLA Survey of the Chandra Deep Field South. \\ IV. Source Population}

\author{P. Padovani, V. Mainieri}
\affil{European Southern Observatory, Karl-Schwarzschild-Str. 2,
D-85748 Garching bei M\"unchen, Germany}
\email{Paolo.Padovani@eso.org}

\author{P. Tozzi}
\affil{INAF, Osservatorio Astronomico di Trieste, Via G. B. Tiepolo 
11, I-34131, Trieste, Italy}

\author{K. I. Kellermann,  E. B. Fomalont}
\affil{National Radio Astronomy Observatory, 520 Edgemont Road, 
 Charlottesville, VA 22903-2475, USA}

\author{N. Miller}

\affil{Department of Physics and Astronomy, Johns Hopkins
University, 3400 North Charles Street, Baltimore, MD 21218, USA}

\author{P. Rosati, P. Shaver}
\affil{European Southern Observatory, Karl-Schwarzschild-Str. 2,
D-85748 Garching bei M\"unchen, Germany}

\begin{abstract}
We present a detailed analysis of 256 radio sources from our deep (flux
density limit of $42~\mu$Jy at the field centre at 1.4 GHz) Chandra Deep
Field South 1.4 and 5 GHz VLA survey. The radio population is studied by
using a wealth of multi-wavelength information in the radio, optical, and
X-ray bands. The availability of redshifts for $\sim 80\%$ of the sources
in our complete sample allows us to derive reliable luminosity estimates
for the majority of the objects. X-ray data, including upper limits, for
all our sources turn out to be a key factor in establishing the nature of
faint radio sources. Due to the faint optical levels probed by this study,
we have uncovered a population of distant Active Galactic Nuclei (AGN)
systematically missing from many previous studies of sub-millijansky radio
source identifications. We find that, while the well-known flattening of
the radio number counts below 1 mJy is mostly due to star forming galaxies,
these sources and AGN make up an approximately equal fraction of the
sub-millijansky sky, contrary to some previous results. The AGN include
radio galaxies, mostly of the low-power, Fanaroff-Riley I type, and a
significant radio-quiet component, which amounts to approximately one fifth
of the total sample. The ratio of radio to optical luminosity depends more
on radio luminosity, rather than being due to optical absorption.
\end{abstract}


%

\section{Introduction}

The extragalactic radio source population ranges from normal galaxies with
luminosities $\sim 10^{20}$ W Hz$^{-1}$ to galaxies whose radio emission is as
much as $10^4 -10^7$ times greater owing to regions of massive star
formation or to an active galactic nucleus (AGN). The population of radio
sources in the sky with flux densities $> 1$ mJy is dominated by AGN driven
emission generated from the gravitational potential associated with a
supermassive black-hole in the nucleus. For these sources, the observed
radio emission includes the classical extended jet and double lobe radio
sources as well as compact radio components more directly associated with
the energy generation and collimation near the central engine. Below 1 mJy
there is an increasing contribution to the radio source population from
synchrotron emission resulting from relativistic plasma ejected from
supernovae associated with massive star formation in galaxies or groups of
galaxies, often associated with mergers or interactions
\citep[e.g.,][]{win95,ric98,fom02}. However, the mix of star forming
galaxies (SFG) and AGN and their dependence on epoch is not well
determined.

For example, \cite{gru03}, by using 15-$\mu$m Infrared Space Observatory
(ISO) and 1.4 GHz data have suggested that starburst galaxies start to
become a significant fraction of the radio source population for $S_{\rm r}
\la 0.5 - 0.8$ mJy, exceeding $\sim 60\%$ only at $S_{\rm r} < 50~
\mu$Jy. On the other hand, \cite{hu05}, by fitting radio number counts of
the Hubble Deep Field South and other surveys with evolutionary starburst
models, have estimated that the number of SFG exceeds
$50\%$ of the total already at $\sim 0.25$ mJy.

The most commonly accepted paradigm has been that the sub-mJy population is
largely made up of SFG. A (small but growing) number of researchers,
however \citep[e.g.,][]{gru99,cil03,sey08,smo08}, have pointed out that
radio sources (radio galaxies and radio-loud AGN) could also make a
significant contribution to the sub-mJy population. It has been further
suggested \citep{jar04,sim06} that radio-quiet AGN (see below) might become
relevant at sub-mJy levels.

The Chandra Deep Field South (CDFS) area is part of the Great Observatories
Origins Deep Survey (GOODS) and as such is one of the most intensely
studied region of the sky. High sensitivity X-ray observations are
available from Chandra \citep{gia02,leh05,luo08}. The GOODS and Galaxy
Evolution from Morphology and Spectral energy distributions (GEMS)
multiband imaging programs using the HST Advanced Camera for Surveys (ACS)
give sensitive high resolution optical images \citep{gia04,rix04}. Ground
based imaging and spectroscopy are available from the ESO 2.2m and 8m
telescopes \citep[see references in][Paper II]{mai08}.

We observed the CDFS with the NRAO Very Large Array (VLA) for 50 hours at
1.4 GHz mostly in the BnA configurations in October 1999 and February 2001,
and for 32 hours at 5 GHz mostly in the C and CnB configurations between
June and October 2001 \citep[details are given in][Paper I]{kel08}. The
effective angular resolution was $3.5''$ and the minimum root-mean-square
(rms) noise was as low as $8.5~\mu$Jy per beam at 1.4 GHz and $7~\mu$Jy per
beam at 5 GHz.  A total of 266 radio sources were catalogued at 1.4 GHz,
198 of which are in a complete sample with signal-to-noise ratio (SNR)
greater than 5 and located within 15$^{\prime}$ of the field center. The
corresponding flux density ranges from 42 $\mu$Jy at the field center to
125 $\mu$Jy near the field edge. These deep radio observations complement
the larger area, but less sensitive, lower resolution observations of the
CDFS discussed by \cite{afo06} and by \cite{nor06}.

In addition to our deep radio data, our set of ancillary data is quite
unique and allows us to shed new light on the nature of the sub-mJy radio
source population. These data include reliable optical/near-IR
identifications for $94\%$
of the radio sources, optical morphological classification for
$61\%$ 
of the sample, redshift information for $73\%$ of the
objects, and X-ray detections for $33\%$ 
of the sources, with upper limits for all but three of the others. This is
the largest and most complete sample of $\mu$Jy sources in terms of
redshift information.

The purpose of this paper is to study the composition of the radio-detected
VLA-CDFS sample in terms of source properties by using all available
information.  More details on the VLA-CDFS can be found in Paper I, which
addresses the radio data, in Paper II, which discusses the optical and near
IR counterparts to the observed radio sources, and in \cite{toz08} (Paper
III), which studies the X-ray properties.

The sample radio-optical data are described in \S~2, while \S~3 deals with
the X-ray data. \S~4 analyzes the overall properties of our sample and
derives the radio number counts for various classes together with some
model predictions, while \S~5 discusses our results. Finally, \S~6
summarizes our conclusions. Throughout this paper spectral indices are
defined by $S_{\nu} \propto \nu^{-\alpha}$ and the values $H_0 = 70$ km
s$^{-1}$ Mpc$^{-1}$, $\Omega_{\rm M} = 0.3$, and $\Omega_{\rm \Lambda} =
0.7$ have been used. This paper supersedes the preliminary results
presented by \cite{pad07b}.

\section{The sample radio-optical properties}\label{sample}

\subsection{Expected radio properties}\label{properties}

Before we turn to the radio sources, we first define the typical ranges in
radio power and radio-to-optical flux density ratio for the type of objects
which are most likely to be part of our survey: galaxies (normal, radio,
and star forming), and active galactic nuclei (AGN).

Normal galaxies, with radio powers $\la 4 \times10^{20}$ W Hz$^{-1}$, are
thought to have their radio emission dominated by synchrotron radiation
from interstellar relativistic electrons \citep{phi86,sad89}. Radio
galaxies are considered to have gigahertz radio power $\ga 10^{22}$ W
Hz$^{-1}$ \citep[e.g.,][]{sad89,led96}, associated with relativistic jets,
kiloparsec scale morphology, which extends well beyond the region of
optical luminosity, and, at least at low redshift, mostly elliptical
hosts. This radio power is at the faint end of the luminosity function of
radio galaxies \citep[e.g.,][]{up95} and, therefore, represents a `natural'
threshold for "radio-loudness" in galaxies, although recent papers have
suggested the presence of non-thermal, jet emission in early type galaxies
as faint as $10^{20}$ W Hz$^{-1}$: e.g., \cite{bal06}.  \cite{fan74}
recognized that radio galaxies separate into two distinct luminosity
classes, each with its own characteristic radio morphology. High-luminosity
Fanaroff-Riley (FR) IIs have radio lobes with prominent hot spots and
bright outer edges, while in low-luminosity FR Is the radio emission is
more diffuse. The luminosity distinction is fairly sharp at 178 MHz, with
FR Is and FR IIs lying below and above, respectively, the fiducial
luminosity $P_{\rm 178 MHz} \approx 10^{25}/(H_0/70)^2$ W Hz$^{-1}$.
At higher radio frequencies the division is at 
$P_{\rm 1.4 GHz} \approx 2 \times 10^{24}/(H_0/70)^2$ W Hz$^{-1}$, with
some dependency also on optical luminosity \citep{ow91} and therefore a
somewhat large overlap. Finally, SFG can also be relatively strong radio
emitters but are hosted by spiral and irregular galaxies at the relative
low redshifts of our sample (see \S~\ref{CDFS-VLA}). These star forming
radio sources dominate the local ($z < 0.4$) radio luminosity function
below $P_{\rm 1.4 GHz} \approx 10^{23}$ W Hz$^{-1}$ but reach only $P_{\rm
  1.4 GHz} \approx 10^{24}$ W Hz$^{-1}$ \citep[e.g.,][]{sad02,be05}, as
compared to $P_{\rm 1.4 GHz} \approx 10^{26}$ W Hz$^{-1}$ or more for the
powerful radio-galaxies. In terms of the rest-frame ratio of radio to
optical flux density, which we define as $R = \log (S_{\rm 1.4GHz}/S_{\rm
  V})$ (where $S_{\rm V}$ is the V-band flux density), SFG have relatively
small values. For example, the majority of the ultra-luminous infrared
galaxies in \cite{veg08} have $S_{\rm 1.4GHz} \approx S_{\rm 1 \mu m}$,
which corresponds to $R \approx 0.4$, while \cite{mac99} have found that
98\% of their normal and starburst galaxies have $R < 1.3$ (converting from
their definition). It is important to note that these two studies reach
only $z \sim 0.1$ and $\sim 0.2$ respectively, and therefore one cannot
exclude the possibility that higher redshift sources might have larger $R$
values (see \S~\ref{CDFS-VLA}). We assume for the time being $R = 1.4$ as
the maximum fiducial value for SFG.  We note that \cite{sey08} have used
the (observed) $S_{\rm 1.4GHz}/S_{\rm 2.2\mu}$ ratio as a discriminant
between AGN and SFG. Although we have $K$ band photometry for less than
half of our sample, we find a very strong correlation between observed
$S_{\rm 1.4GHz}/S_{\rm 2.2\mu}$ and rest-frame $S_{\rm 1.4GHz}/S_{\rm V}$,
from which we infer that Seymour et al.'s dividing line corresponds to $R =
1.4\pm0.1$, consistent with our own (but see \S~\ref{population}).

About $\sim 10\%$ of AGN have, for the same optical power, radio powers $3
- 4$ orders of magnitude higher than the rest. Two different definitions of
radio-loudness have been used in the literature \citep[see,
  e.g.,][]{kel89,sto92}. The first is based on the rest-frame ratio of the
(5 GHz) radio to optical flux density, with radio-loud sources having
values $\ga 10$. Assuming $\alpha_{\rm r} = 0.7$ (see below), and
converting to 1.4 GHz, the commonly accepted dividing line between
radio-loud and radio-quiet AGN translates then to $R \sim 1.4$, which is
the same as the maximum value for SFG adopted above. The other definition
is based on gigahertz radio power and appears to be redshift-dependent,
with a dividing line ranging from $\approx 5 \times 10^{23}$ W Hz$^{-1}$
for the Palomar Green sample (optically bright and at relatively
low-redshift) up to $5 \times 10^{25}$ W Hz$^{-1}$ for the Large Bright
Quasar Survey sample (optically fainter and at higher redshift). The two
definitions overlap somewhat \citep{pad93}. The radio power limit is, not
surprisingly, very close to the maximum power which can be reached by
SFG. In the following we will use a single value, $P_{\rm 1.4 GHz} \approx
10^{24}$ W Hz$^{-1}$, appropriate for our relatively low redshift sample
(see below), and comment also on how our results depend on this value.

\subsection{The CDFS-VLA sample}\label{CDFS-VLA}

Our sample includes all CDFS-VLA sources with a highly reliable (likelihood ratio 
$> 0.2$ as defined in Paper II) optical counterpart (apart from
one source very close to a bright star for which we could not get reliable
photometry) and eight empty fields, for a total of 256 objects, 194 of
which belong to the complete sample. We include all sources when studying
the overall sample properties but we obviously restrict ourselves to the
complete sample when dealing with number counts and fractions of sources of
a given class.

Our usage of $V$ band data to derive the radio-to-optical flux density
ratio is dictated by two facts: 1. almost complete coverage for our sources
(we estimated $V$ magnitudes from other bands for 20 sources using mean
colors); 2. consistency with previous definitions of radio-loud AGN. However,
our results are broadly independent of this choice.

One hundred and eight ($42\%$) sources have spectroscopic redshifts,
while photometric redshifts are also available for 78 ($31\%$) objects
(see details in Paper II).  In total, we have redshift information for
$73\%$ of our objects, with this fraction reaching $77\%$ for the
complete sample. The distribution peaks at $z \sim 0.7$, with a sharp
cut-off at $z > 1$ (see Fig. 11 in Paper II). The mean redshift is $\langle
z \rangle = 0.80\pm0.05$ (median 0.67).

\begin{figure}
\centerline{\includegraphics[width=9.5cm]{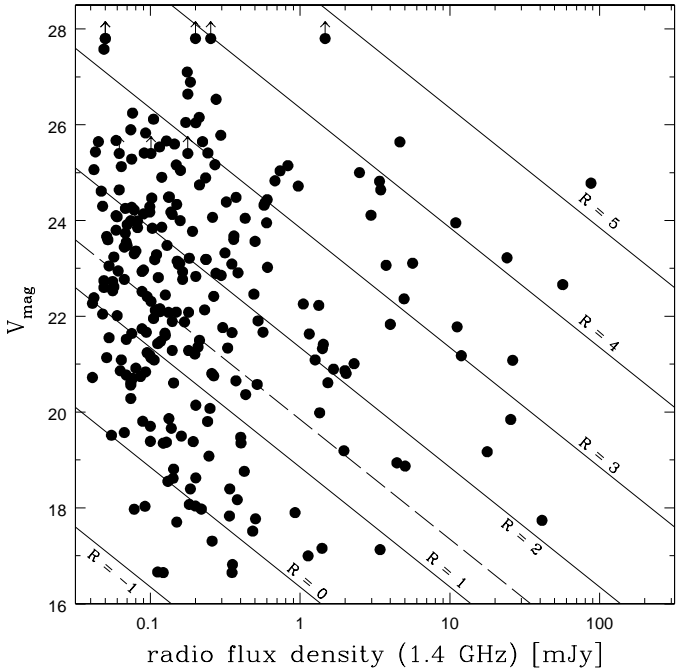}}
\caption{The $V$ band magnitude versus the 1.4 GHz radio flux density for
  the CDFS-VLA sample. Arrows represent lower limits of $V_{\rm mag}$. The
  lines represent different values of the logarithm of the observed
  radio-to-optical flux density ratio $R = \log(S_{\rm 1.4GHz}/S_{\rm V})$,
  ranging from 5 (top right) down to $-1$ (bottom left). The dashed line
  indicates the maximum local value for SFG and the approximate dividing
  line between radio-loud and radio-quiet AGN ($R =
  1.4$). } \label{vfrcdfs}
\end{figure}

\begin{figure}
\centerline{\includegraphics[width=9.5cm]{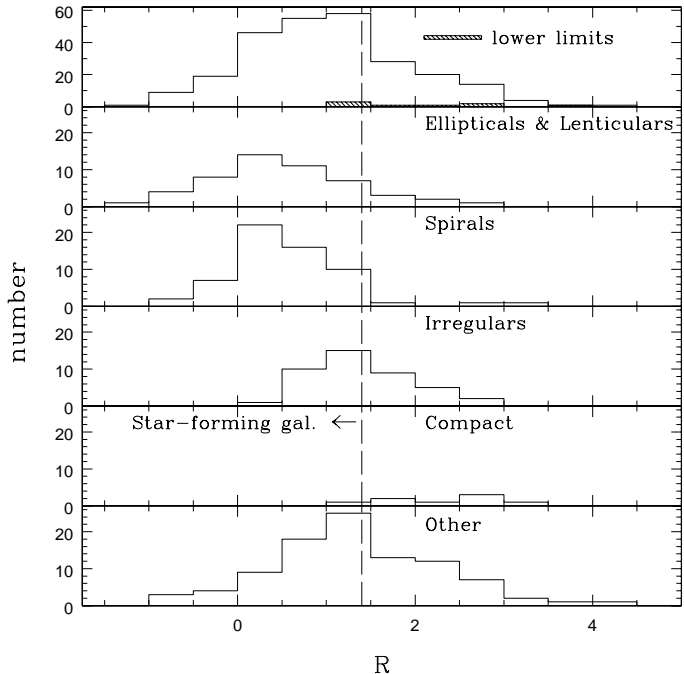}}
\caption{The distribution of the rest-frame radio-to-optical flux density
  ratio $R = \log (S_{\rm 1.4GHz}/S_{\rm V})$. {\it Top panel:} Full
  sample. The dashed area represents lower limits. The
  vertical dashed line indicates the approximate maximum value for local
  SFG and dividing line between radio-loud and radio-quiet AGN. {\it Other
    panels:} Objects sorted by morphological classes, namely, from top to
  bottom: ellipticals and lenticulars, spirals, irregulars, compact, and
  sources which are either not easily classifiable, faint, or complex.}
\label{histr}
\end{figure}

Fig. \ref{vfrcdfs} shows the $V$ band magnitude versus the 1.4 GHz radio
flux density for the CDFS-VLA sample. The lines represent different values
of the logarithm of the radio-to-optical flux density ratio, ranging from 5
(top right) down to $-1$ (bottom left). We note that $63\%$ of the
sources have observed $R$ values beyond the star forming regime, with some
of the sources reaching extremely high values of $R$, equivalent to
radio-to-optical flux density ratios $\approx 10^5$.

This is better seen in Fig. \ref{histr} (top panel), which shows the
distribution of rest-frame $R$. For K-correction purposes we used the
observed radio spectral index, when available, and the mean value, $\langle
\alpha_{\rm r} \rangle \simeq 0.7$, when that was not the case. In the
optical band K-correction values appropriate for the morphological type of
the objects were adopted, based on \cite{col80}. Sources without
morphological classification were assumed to be ellipticals\footnote{Note
  that, given the larger K-corrections of ellipticals, any other assumption
  would result in {\it larger} $R$ values.}  (see below), while for compact
objects (likely AGN), $\alpha_{\rm o} = 0.5$ was assumed. The mean value is
$\langle R \rangle = 1.04\pm0.06$. The distribution of $R$ for the objects
without redshift (derived assuming $z = \langle z \rangle$) is widely
different ($P > 99.99\%$ according to a Kolmogorov-Smirnov test) from that
of the remaining sources. The mean values for the two sub-samples are in
fact $\langle R \rangle = 1.62\pm0.12$ and $\langle R \rangle =
0.83\pm0.06$ respectively, with $60\%$ of the sources without redshift
being in the star forming regime. Their larger $R$ values are due to
their fainter $V$ magnitudes, as the radio flux density distributions for
the two sub-samples are not significantly different.

Fig. \ref{histr} (lower panels) shows the $R$ distribution for various
morphological classes (see Paper II for more details), namely ellipticals
and lenticulars (51), spirals (60), irregular galaxies (including mergers;
42), compact objects (8), and sources which are either not easily
classifiable, too faint, or complex (95).
The various types have somewhat different $R$ values, as shown in
Table \ref{tab_mean}. Compact sources have the largest $\langle R \rangle$
value, followed by irregulars, and then galaxies which are not easily
classifiable. Ellipticals/lenticulars and spirals have the lowest
radio-to-optical flux density ratios. The $R$ distribution for galaxies not
classifiable is significantly different ($P \sim 98.5\%$) from that of
spiral/irregular galaxies and also from that of ellipticals/lenticulars ($P
> 99.99\%$). Somewhat surprisingly, spirals and irregulars, which we can
group together as likely hosts of SFG, have $R$ larger
than ellipticals/lenticulars, with the two $R$ distributions being
significantly different ($P \sim 98.8\%$). Only $12\%$ of elliptical/lenticular galaxies have $R
> 1.4$, while $21\%$ of spirals and irregulars are above this value. 
We find a significant difference also between spirals and irregulars ($P >
99.99\%$).
It has been suggested that the radio-to-optical flux density ratio alone
cannot be used to discriminate between AGN and SFG
\citep[e.g.,][]{afo05,bar07}. We confirm that this is indeed the case. 

Possible causes for the fact that spirals and irregulars in our sample
reach relatively high radio-to-optical flux density ratios, include: a) $R$
evolution. Our sample has $\langle z \rangle \sim 0.8$, reaching $z \sim
2$, as compared to the local star forming samples for which $R$ has been
previously derived. We have quantified possible evolutionary effects by
assuming a radio evolution $\propto (1+z)^{2.7}$ \citep{hop04} and passive
$V$-band evolution for Sa and Sc galaxies \citep{col80}. In the Sa case $R$
is predicted to {\it decrease} at higher redshifts, because the optical
evolution is stronger than the radio one. For Sc galaxies, on the other
hand, there is a weak evolution with $R(z) \proptoapprox \log (1+z)$. This
means that at $z = \langle z \rangle$ $R_{\rm max} \la 1.7$ and at $z \sim
2$ $R_{\rm max}$ should be $\la 1.9$. This effect could explain up to half
of the $R > 1.4$ sources with redshift; b) uncertainties in morphological
classification. While Paper II has shown that overall our morphological
classification is robust, there could still be subtle effects, which might
introduce some bias. For example, while the distributions of magnitudes of
ellipticals/lenticulars and spiral galaxies in our sample are very similar,
irregulars are on average two magnitudes fainter. If some galaxies are
preferentially classified as such when they are faint, this will result in
artificially large $R$ values for irregulars, with the effect being
amplified by the fact that the K-correction for irregulars is the smallest
one; c) K-correction. We assumed standard K-corrections, which imply that
all spectral energy distributions (SEDs) for sources of a given class are
the same. This is obviously unrealistic and will certainly introduce a
scatter in the $R$ distribution; d) the presence of sources like the one
discovered by \cite{nor08}: a classic triple radio source at $z = 0.93$ and
$P_{\rm r} \sim 4 \times 10^{25}$ W Hz$^{-1}$ with the SED of a spiral
galaxy. Finally, it is important to notice that not all spirals and
irregulars are SFG, as potentially many of these sources could be AGN. It
is then clear that we need to add luminosity information.

\begin{deluxetable*}{lcrcrc}
\tabletypesize{\scriptsize}
\tablecaption{Mean values. \label{tab_mean}}
\tablewidth{0pt}
\tablehead{Galaxy class& $\langle R \rangle$ & N ~~~~& $\langle \log P_r
  \rangle$ & N ~~~~& $\langle \log L_x \rangle$ \\}
\startdata
Ellipticals and lenticulars & $0.52\pm0.11$ & 51~~~~& $23.47\pm 0.16$ &
47~~~~& $41.55\pm 0.23$\\
Spirals & $0.58\pm0.09$ & 60~~~~& $22.75\pm 0.12$ & 59~~~~& $40.65\pm 0.16$\\
Irregulars & $1.42\pm0.09$ & 42~~~~& $23.38\pm 0.11$ & 36~~~~& $41.49\pm 0.27$\\
Spirals and Irregulars & $0.92\pm0.07$ & 102~~~~& $22.99\pm0.09$ & 95~~~~&
$40.87\pm0.15$\\
Compact & $2.40\pm0.27$ & 8~~~~& $24.62\pm 0.66$ & 6~~~~& $43.54\pm 0.55$\\
Other & $1.34\pm0.10$ & 95~~~~& $24.05\pm 0.13$ & 38~~~~& $42.55\pm 0.16$\\
No redshift, $z = \langle z \rangle$ & $1.62\pm0.12$ & 70~~~~& $23.92\pm
0.09$ & 70~~~~& $41.85\pm 0.13$\\
No redshift, $z = 1.0$ & $1.31\pm0.12$ & 70~~~~& $24.15\pm 0.09$ & 70~~~~&
$42.06\pm 0.13$\\
No redshift, $z = 1.5$ & $0.83\pm0.13$ & 70~~~~& $24.56\pm 0.09$ & 70~~~~&
$42.41\pm 0.13$\\
No redshift, $z = 2.0$ & $0.68\pm0.14$ & 70~~~~& $24.85\pm 0.09$ & 70~~~~&
$42.68\pm 0.13$\\
\enddata
\end{deluxetable*}

\begin{figure}
\centerline{\includegraphics[width=9.5cm]{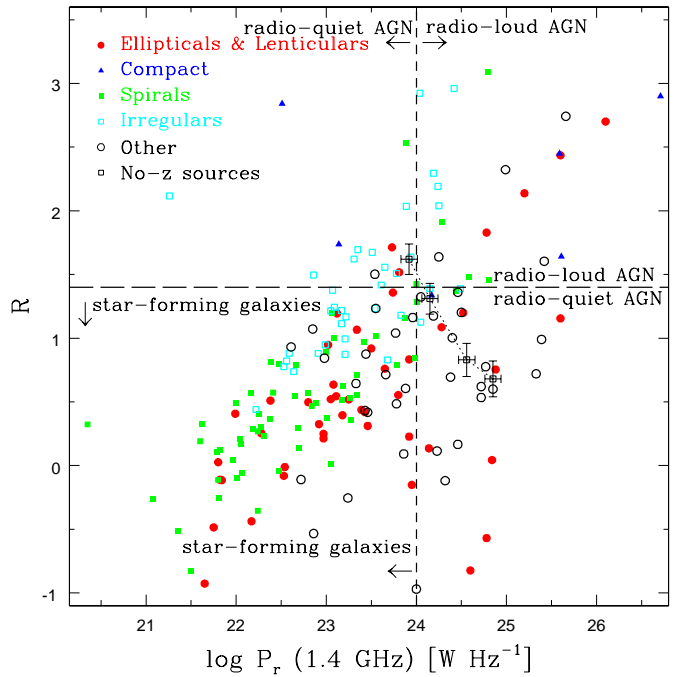}}
\caption{The rest-frame radio-to-optical flux density ratio $R = \log
  (S_{\rm 1.4GHz}/S_{\rm V})$ versus the 1.4 GHz radio power for the
  sources with redshift information. The horizontal (long-dashed) line at
  $R = 1.4$ and the vertical (short-dashed) line at $\log P_{\rm r} = 24$
  indicate the approximate maximum values for local SFG and dividing lines
  between radio-loud and radio-quiet AGN respectively.
  Sources are classified morphologically, namely: ellipticals and
  lenticulars, red filled circles; compact, blue triangles; spirals, green
  filled squares; irregulars, cyan open squares; sources which are either
  galaxies not easily classifiable, faint, or complex, black open
  circles. The black open squares with error brackets represent the locus
  of objects without redshift assuming (from left to right) $z = \langle z
  \rangle$, $z=1.0$, $z=1.5$, and $z=2$ and an optical K-correction
  appropriate for ellipticals. See text for more details.}  \label{rloglr}
\end{figure}

Fig. \ref{rloglr} shows $R$ versus the 1.4 GHz radio power for the sources
with redshift information.  Apart from the obvious correlation between $R$
and $P_{\rm r}$ (the former being equal to $P_{\rm r}/P_{\rm opt}$),
Fig. \ref{rloglr} shows that
all but three sources with $R > 1.4$ have $P_{\rm r} \ga 10^{23}$ W Hz$^{-1}$.
Turning to morphological source classification, the following points can be
made: a) the large majority (85\%) of spirals and irregulars have $P_{\rm
  r} \le 10^{24}$ W Hz$^{-1}$, with this fraction reaching $97\%$ and 100\% for
$P_{\rm r} \le 3 \times 10^{24}$ W Hz$^{-1}$ and $P_{\rm r} \le 10^{25}$ W Hz$^{-1}$
respectively; b) spirals and irregulars with $R > 1.4$ have relatively high
radio powers ($\langle P_{\rm r} \rangle \sim 6 \times 10^{23}$ W Hz$^{-1}$) and
basically all of the sources with $R > 2$ are above (or very close to) the
$10^{24}$ W Hz$^{-1}$ limit; c) 5/6 of compact sources are consistent with being
radio-loud AGN; d) ellipticals and lenticulars span a very large range of
radio powers, consistent with that of radio-galaxies; there is substantial
overlap at low powers between early- and late-type galaxies, with 75\% of
the former having $P_{\rm r} \le 10^{24}$ W Hz$^{-1}$; e) all "other" sources have
powers in the range of radio galaxies and half of them actually reach into
the radio-loud AGN regime. Most of the various types have quite distinct
radio powers, as shown in Table \ref{tab_mean}. Compact sources have the
largest radio powers, followed by galaxies which are not easily
classifiable, and ellipticals/lenticulars. Spirals/irregulars have the
lowest values. The radio power distribution for the sources without
morphological classification is significantly different ($P > 99.99\%$)
from that of spirals and irregulars and also ($P \sim 98.8\%$) from that of
early-type galaxies.

We have also optical spectral classification, from \cite{szo04}, for a
relatively small fraction ($14\%$) of the sample, which we use to
gather further information. This includes Broad Line AGN (BLAGN), high
excitation line objects (HEX), low excitation line objects (LEX), Narrow
Emission Line Galaxies (NELG), and objects with normal galaxy spectra
(ABS).  Broadly speaking, BLAGN correspond to type-1 AGN, HEX mostly to
type-2 AGN, while the properties of NELG, LEX, and ABS are consistent with
those of normal galaxies but in many cases this is simply a limitation of
the optical data.

The spectral classification adds the following information: a) AGN have, on
average, radio powers larger than those of LEX, ABS, and NELG ($\sim
10^{24}$ W Hz$^{-1}$ as compared to $\sim 8 \times 10^{22}$ W Hz$^{-1}$);
b) all but one LEX, ABS, and NELG have $P_{\rm r} < 10^{24}$ W Hz$^{-1}$
and $\sim 78\%$ of them concentrate in the bottom-left quadrant of
Fig. \ref{rloglr}, consistent with the fact that this is where SFG are
located.

As for the radio-to-optical flux density ratio, evolution will also affect
the maximum radio power of SFG. For $P_{\rm r} \propto (1+z)^{2.7}$
\citep{hop04}, this could reach $P_{\rm r} \sim 5 \times 10^{24}$ W
Hz$^{-1}$ and $P_{\rm r} \sim 2 \times 10^{25}$ W Hz$^{-1}$ at $z = \langle
z \rangle$ and $z \sim 2$ respectively.

Fig. \ref{rloglr} does not include sources without redshift. If we assign a
redshift equal to the mean value of the sample to such sources, they would
have $\langle R \rangle \sim 1.6$ and $\log P_{\rm r} \sim 23.9$
(Table \ref{tab_mean}), that is most of them would fall into a region
outside that of SFG (see Fig. \ref{rloglr}). For $z = 1 - 2$, which are
more likely values given the dependence we find between redshift and
magnitude, radio powers would obviously increase, with $R$ values
decreasing due to the larger K-correction (as $\sim 80\%$ of such sources
have no morphological information and we assume that they are ellipticals),
as shown in Fig. \ref{rloglr}. However, the point remains that the majority
of these sources would not have values very typical of SFG.

In summary, $R$ is not a very good discriminant between SFG and AGN. Radio
power fares somewhat better but the overlap between the two classes is
still quite large. This is obviously particularly bad for the sources for
which no optical morphology is available. 

\section{Adding X-ray information}\label{xrayinfo}

\begin{figure}
\centerline{\includegraphics[width=9.5cm]{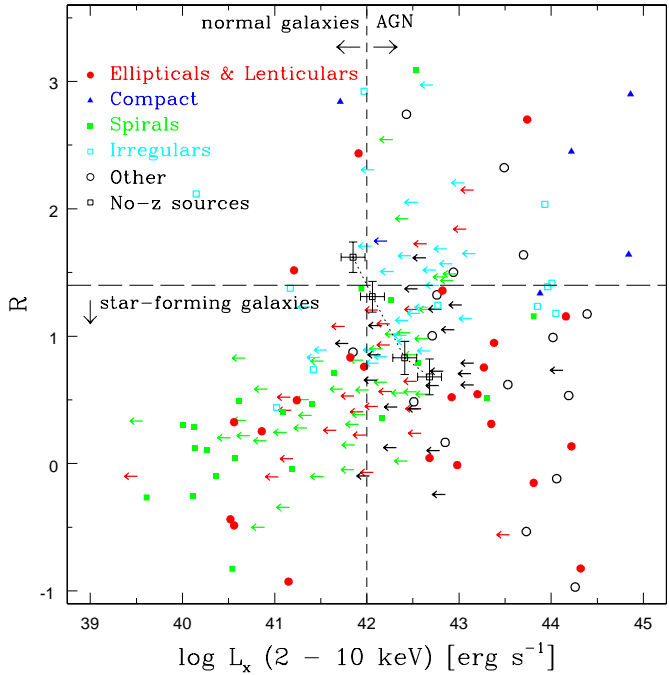}}
\caption{The rest-frame radio-to-optical flux density ratio $R = \log
  (S_{\rm 1.4GHz}/S_{\rm V})$ versus the $2 - 10$ keV X-ray power for the
  sources with redshift information. The horizontal (long-dashed) line at
  $R = 1.4$ indicates the approximate maximum value for local SFG and
  dividing line between radio-loud and radio-quiet AGN, while the vertical
  (short-dashed) line at $L_{\rm x} = 10^{42}$ ergs s$^{-1}$ indicates the
  dividing line between AGN and normal galaxies. Sources are classified
  morphologically, namely: ellipticals and lenticulars, red filled circles;
  compact, blue triangles; spirals, green filled squares; irregulars, cyan
  open squares; sources which are either galaxies not easily classifiable,
  faint, or complex, black open circles. The black open squares with error
  brackets represent the locus of objects without redshift assuming (from
  left to right) $z = \langle z \rangle$, $z=1.0$, $z=1.5$, and $z=2$ and
  $z=2$ and an optical K-correction appropriate for ellipticals.  See text
  for more details.}  \label{rloglx}
\end{figure}

We now consider the X-ray information to learn more about the nature of the
radio sources.  Fig. \ref{rloglx} plots the rest-frame radio-to-optical
flux density ratio $R$ vs. the $2 - 10$ keV power for the sources with
redshift information. X-ray powers and upper limits are available for all
but three sources and for all of those in the complete sample (see Paper
III). The vertical short-dashed line at $L_{\rm x} = 10^{42}$ ergs s$^{-1}$
indicates the approximate dividing line between AGN and normal galaxies
\citep[e.g.,][]{szo04}.

To properly take into account the upper limits on X-ray power we used ASURV
\citep{la92}, the Survival Analysis package which employs the routines
described in \cite{fei85} and \cite{iso86}, which evaluate mean values by
dealing properly with non-detections. Survival analysis methods also
evaluate probability of correlation and linear regression fits by dealing
properly with non-detections. A correlation found between luminosities,
when such non-detections are included, is a true correlation, while flux
density - flux density diagrams, which are sometimes used, will reveal the
intrinsic relationship between luminosities only under very limited
circumstances (no censored points, linear relationship). 
Since upper limits make up the majority ($\sim 67\%$) of the sample,
the uncertainties on our results are likely to be larger than the ones
given by the software package.

We find a very strong ($P > 99.0\%$) correlation between $L_{\rm x}$ and R
for the whole sample, which is even stronger ($P > 99.9\%$) for spirals and
irregulars. Two striking features of Fig. \ref{rloglx} are: a) the fact
that $58\%$ of all X-ray detections have $L_{\rm x} > 10^{42}$ ergs s$^{-1}$,
with this fraction reaching $62\%$ if one includes upper limits as
well; b) the almost complete lack of objects in the top left quadrant, as
only $3\%$ of the sources have $R > 1.4$ and $L_{\rm x} < 10^{42}$
ergs s$^{-1}$ (only half of those are irregulars). Indeed, $R \ge 1.4$
sources have $\langle \log L_{\rm x} \rangle = 42.14\pm0.28$, almost an
order of magnitude larger than $R < 1.4$ sources, which have $\langle \log
L_{\rm x} \rangle = 41.21\pm0.13$.
This suggests that a large fraction of our sources are AGN and that many high 
$R$ sources with spiral/irregular morphology {\it cannot} be SFG 
as their X-ray powers are too high. 

As was the case for the radio power, most of the various types have quite
distinct X-ray powers, as shown in Table \ref{tab_mean}.
Compact sources have the largest X-ray powers, followed closely by galaxies
which are not easily classifiable, and ellipticals/lenticulars.
Irregulars/spirals have the lowest values. The X-ray power distribution for
spirals and irregulars is significantly different\footnote{The ASURV
  package evaluates the difference between two distribution using five
  different tests. We quote here the {\it lowest} probability.} from that
of the sources without morphological classification ($P > 99.98\%$) and
also from that of early-type galaxies ($P > 98.6\%$). The spectral
classification adds the fact that AGN, on average, have X-ray powers larger
than those of LEX, ABS, and NELG ($\sim 2 \times 10^{43}$ ergs s$^{-1}$ as
compared to $6 \times 10^{41}$ ergs s$^{-1}$), although $\sim 30\%$ of LEX
and ABS sources are in the AGN region (all BLAGN and all but but two HEX
have AGN-like X-ray powers); in particular, all NELG have only upper limits
on their X-ray fluxes.

As before, Fig. \ref{rloglx} does not include sources without redshift. If
we assign a redshift equal to the mean value of the sample to such sources,
they would have $\langle R \rangle \sim 1.6$ and $\log L_{\rm x} \sim 41.8$
(Table \ref{tab_mean}), that is many of them would fall into a region
outside that of SFG. For $z = 1 - 2$, which are more likely values given
the dependence we find between redshift and source magnitude, $R$ would
decrease but X-ray powers would obviously increase, as shown in
Fig. \ref{rloglx}, with the result that the majority of these sources would
not have values typical of SFG.

\begin{figure}
\centerline{\includegraphics[width=9.5cm]{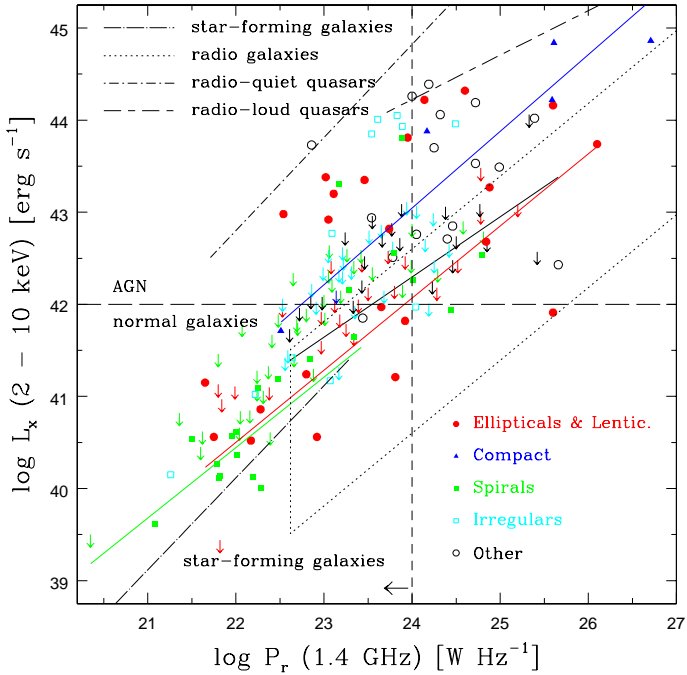}}
\caption{The $2 - 10$ keV X-ray power vs. the 1.4 GHz radio power for
  sources with redshift and X-ray information. The horizontal (long-dashed)
  line at $L_{\rm x} = 10^{42}$ ergs s$^{-1}$ indicates the dividing line between
  AGN and normal galaxies, while the vertical (short-dashed) line at
  $P_{\rm r} = 10^{24}$ W Hz$^{-1}$ shows the maximum $P_{\rm r}$ of SFG at $z
  \sim 0$, which coincides with the dividing lines between low-$z$
  radio-loud and radio-quiet AGN. The dot-long-dashed line is the $L_{\rm
    x} - P_{\rm r}$ relationship for nearby SFG \citep{ra03}, converted to
  our cosmology, the dotted lines denote the locus of radio galaxies in the
  AGN catalogue of \cite{pad97}, while the dot-short-dashed and
  short-long-dashed lines are the best fits for radio-quiet and radio-loud
  quasars from \cite{bri00}. Arrows represent X-ray upper limits. Sources
  are classified morphologically, namely: ellipticals and lenticulars, red
  filled circles; compact, blue triangles; spirals, green filled squares;
  irregulars, cyan open squares; sources which are either galaxies not
  easily classifiable, faint, or complex, black open circles. Solid lines
  show the best fits in each morphological category taking the upper limits
  into account (up to $\log P_{\rm r} < 23.5$ and excluding X-ray
  detections with $\log L_{\rm x} > 42$ for spirals/irregulars: green
  line).}
\label{loglrloglx} \end{figure}

Fig. \ref{loglrloglx} plots $2 - 10$ KeV X-ray power versus the 1.4 GHz
radio power for the sources with redshift information. The horizontal,
long-dashed line at $L_{\rm x} = 10^{42}$ ergs s$^{-1}$ indicates the
approximate dividing line between AGN and normal galaxies
\citep[e.g.,][]{szo04}, while the vertical, short-dashed line at $\log
P_{\rm r} = 24$ indicates the approximate maximum radio power (at $z \sim
0$) due to star-formation, which roughly coincides with the dividing lines
between low-redshift radio-loud and radio-quiet AGN. The dot-long-dashed
line is the X-ray-radio power relation for nearby SFG \citep{ra03}, the
dotted lines represent the approximate locus of radio galaxies in the AGN
catalogue of \cite{pad97}, while the dot-short-dashed and short-long-dashed
lines are the best fits for radio-quiet and radio-loud quasars from
\cite{bri00}, where their 2 keV powers were converted to the $2 - 10$ keV
band assuming $\alpha_{\rm x} = 0.8$. For K-correction purposes we used the
observed radio and X-ray spectral indices, when available, and the mean
values, $\langle \alpha_{\rm r} \rangle \simeq 0.7$ and $\langle
\alpha_{\rm x} \rangle \simeq 0.8$, when that was not the case.

Fig. \ref{loglrloglx} shows that there is a correlation between radio and
X-ray luminosity for the whole sample. ASURV confirms this with a very high
significance ($P > 99.99\%$) giving $L_{\rm x} \propto P_{\rm
  r}^{0.93\pm0.08}$.  Basically all sources in the radio-loud AGN region
have AGN-like X-ray powers. The opposite, however, is not true, that is not
all sources with $L_{\rm x} > 10^{42}$ ergs s$^{-1}$ have $P_{\rm r} \ga
10^{24}$ W Hz$^{-1}$. This makes sense, as not all AGN are
radio-loud. Turning to morphological source classification, the following
points can be made: a) 38\% of spiral and irregular galaxies have X-ray
detections inconsistent with the powers typical of galaxies ($\ga 10^{42}$
ergs s$^{-1}$); that is, a sizable fraction of these sources are most
likely powered by AGN. The best fit correlation using ASURV (green line)
excluding sources with $P_{\rm r} > 10^{23.5}$ W Hz$^{-1}$ for consistency
with \cite{ra03} and X-ray detections with $L_{\rm x} > 10^{42}$ ergs
s$^{-1}$ to select against AGN, is $L_{\rm x} \propto P_{\rm
  r}^{0.8\pm0.1}$ ($P \ge 99.6\%$).  Given the relatively large fraction of
X-ray upper limits ($\sim 70\%$), we regard this as not inconsistent with
the X-ray-radio power relationship for nearby SFG (dot-dashed line); b) as
was the case for the radio power, ellipticals and lenticulars span a very
large range of X-ray powers, consistent with that of radio-galaxies (red
line); again, there is substantial overlap at low X-ray powers between
ellipticals and spirals; c) "other" sources have mostly $L_{\rm x} \ga
10^{42}$ ergs s$^{-1}$, with the best fit correlation (black line) into the
radio-galaxies area; d) compact sources have X-ray powers $ > 5 \times
10^{41}$ ergs s$^{-1}$ and $\approx 10$ times larger than the other classes
at a given radio power.

Note that the correlation we find for spirals/irregulars disagrees with
\cite{bar07}, who suggested that the Ranalli et al. correlation was due to
selection effects, but agrees with other results (see discussion in Paper
III).

We also notice that only a relatively small number of sources cluster
around the relationships previously derived for radio-quiet and radio-loud
quasars. We will return to this point later in \S~\ref{population}. Here we
point out that the fraction of QSOs, defined as sources with $L_{\rm x} >
10^{44}$ ergs s$^{-1}$, is only $7\%$ of the objects with redshift and
X-ray information. Note that this has to be considered a robust upper
limit, as radio galaxies are also strong X-ray emitters. For example, the
relatively low-redshift ($\langle z \rangle = 0.1$) radio galaxies in the
AGN catalogue of \cite{pad97} can reach $L_{\rm x} \sim 3 \times 10^{44}$
ergs s$^{-1}$, and {\it all} five 3CR radio galaxies studied by
\cite{sal08} ($\langle z \rangle = 1.3$) have $L_{\rm x} > 10^{44}$ ergs
s$^{-1}$. BL Lacs are even stronger X-ray emitters, with $\sim 70\%$ of
those in the AGN catalogue of \cite{pad97} having $L_{\rm x} > 10^{44}$
ergs s$^{-1}$, but are likely to be a minority population in our sample
(see \S~\ref{expected}). Indeed, the average radio spectral index for the
$L_{\rm x} > 10^{44}$ ergs s$^{-1}$ sources is $\sim 0.6$,
with only $\sim 20\%$ of them having $\alpha_{\rm r} < 0.5$ and therefore
qualifying as flat-spectrum, which is one of the defining properties of BL
Lacs.

In summary, the X-ray data appear to be quite powerful in recognizing
AGN
among the sub-mJy population.

\section{The sub-mJy source population}\label{population}

\subsection{The redshift sub-sample}\label{redshift}

We now examine collectively the radio, optical, and X-ray data. We start by
considering only sources with redshift information, which include $77\%$ 
of the complete sample. Based on the results presented in the
previous sections, our candidate star-forming galaxies are defined as
fulfilling all the following requirements:

\begin{enumerate}

\item $R < 1.7$

\item $P_{\rm r} < 10^{24.5}$ W Hz$^{-1}$

\item optical morphology different from elliptical or lenticular

\item $L_{\rm x} < 10^{42}$ ergs s$^{-1}$ for X-ray detections, no limit
  otherwise.

\end{enumerate}

The first two criteria include $87\%$ of spirals and irregulars, with
the remaining fraction being most likely AGN, the third one excludes
sources not associated with star formation at the redshifts under
consideration, while the fourth one excludes AGN.

\begin{figure}
\centerline{\includegraphics[width=9.5cm]{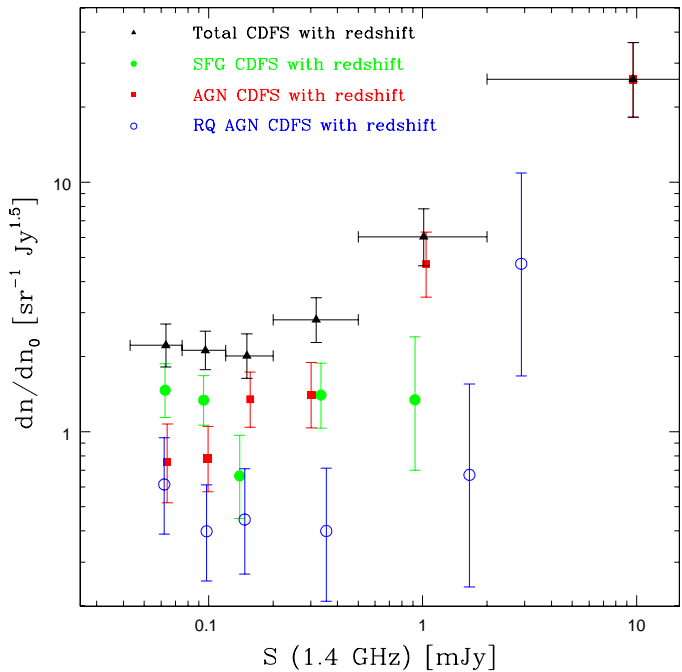}}
\caption{The Euclidean normalized 1.4 GHz CDFS source counts for sources
  with redshift information: total counts (black triangles), star forming
  galaxies (filled green circles), all AGN (red squares), and radio-quiet
  AGN (open blue circles).  Error bars correspond to $1\sigma$ errors
  \citep{geh86}.} \label{counts_z}
\end{figure}

\begin{deluxetable*}{lrccrcr}
\tabletypesize{\scriptsize}
\tablecaption{Euclidean normalized 1.4 GHz counts for sources with
  redshift. \label{tab_counts_z}}
\tablewidth{0pt}
\tablehead{Flux range & Mean Flux Density &   & & Counts &  & \\
                 & & ##Total & SF & fraction & AGN & fraction \\
 ###$\mu$Jy & $\mu$Jy & sr$^{-1}$ Jy$^{1.5}$ & sr$^{-1}$ Jy$^{1.5}$ & \% &
  sr$^{-1}$ Jy$^{1.5}$ & \% \\
}
\startdata
$##43 - 75$ & #63 &$#2.22^{+0.48}_{-0.40}$ & $1.47^{+0.41}_{-0.32}$ &
$66^{+22}_{-20}$ & $0.75^{+0.32}_{-0.23}$ & $34^{+16}_{-13}$ \\
$##75 - 120$ & #97 &$#2.12^{+ 0.41}_{-0.35}$ & $1.34^{+0.34}_{-0.28}$ &
$63^{+19}_{-18}$ & $0.78^{+0.27}_{-0.21}$ & $37^{+14}_{-12}$ \\
$#120 - 200$ & 151 & $#2.01^{+0.46}_{-0.38}$ & $0.67^{+0.30}_{-0.22}$&
$33^{+16}_{-13}$ & $1.35^{+0.39}_{-0.31}$ & $67^{+23}_{-22}$\\
$#200 - 500$ & 318 & $#2.81^{+0.64}_{-0.53}$ & $1.40^{+0.48}_{-0.37}$&
$49^{+19}_{-17}$ & $1.41^{+0.49}_{-0.37}$ & $50^{+20}_{-17}$ \\
$#500 - 2000$ & 1011 & $#6.04^{+1.78}_{-1.41}$ & $1.34^{+1.06}_{-0.64}$&
$22^{+18}_{-12}$ & $4.70^{+1.62}_{-1.24}$ & $78^{+32}_{-31}$\\
$2000 - 100000$ & 9639 &$25.9^{+10.4}_{-7.7}$ & 0.0& 0## 
&$25.9^{+10.4}_{-7.7}$ & $100^{+50}_{-50}$\\
\enddata
\end{deluxetable*}

Objects not fulfilling these criteria were considered to be AGN. Note that
the last requirement is conservative in the sense that it maximizes the
number of SFG, as some of the sources with X-ray upper limits above the
X-ray threshold could still have $L_{\rm x} > 10^{42}$ ergs s$^{-1}$ (see below).

Radio-quiet AGN were also selected by considering sources with $R < 1.4$
and X-ray detections with $L_{\rm x} > 10^{42}$ ergs s$^{-1}$. We did not
apply any cut in radio power since the objects with $P_{\rm r} > 10^{24}$ W
Hz$^{-1}$ are all at relatively high redshifts ($\langle z \rangle \sim 2$)
and are therefore expected to have relatively large luminosities even if
radio-quiet (see \S~\ref{properties}).

Table \ref{tab_counts_z} and Fig. \ref{counts_z} present the Euclidean
normalized number counts for the total redshift sub-sample and also for SFG
and AGN, derived as described in Paper I. AGN make up $41^{+7}_{-6}\%$ of
sub-mJy sources with redshift information but their number drops towards
lower flux densities, going from 100\% at $\sim 10$ mJy down to $34\%$
at the survey limit. SFG, on the other hand, constitute $59^{+10}_{-9}\%$
of the sub-mJy sample having measured redshifts. SFG are missing at high
flux densities but become the dominant population below $\approx 0.1$ mJy,
reaching $66\%$ at the survey limit. Radio-quiet AGN represent
$23^{+6}_{-5}\%$ (or $\sim 56\%$ of all AGN) of sub-mJy sources but their
fraction appears to increase at lower flux densities, where they make up
$81\%$ of all AGN at the survey limit.

\begin{deluxetable}{rllll}
\tabletypesize{\scriptsize}
\tablecaption{Fraction of sub-mJy star forming galaxies for various
  parameter limits. \label{perturb}}
\tablewidth{0pt}
\tablehead{R & $\log P_{\rm r}$ &  $\log L_{\rm x}$  & N &  sub-mJy fraction \\
     & W Hz$^{-1}$                   & ergs s$^{-1}$                       &                                       &  \% \\}
\startdata
$< 1.7$ & $< 24.5$ & $< 42$ & 70 & $59^{+10}_{-9}$ \\
$< 1.4$ & $< 24.5$ & $< 42$ & 64 & $53^{+9}_{-8}$ \\
$< 2.0$ & $< 24.5$ & $< 42$ & 72 & $61^{+10}_{-9}$ \\
$< 2.3$ & $< 24.5$ & $< 42$ & 76 & $62^{+10}_{-9}$ \\
$< 1.7$ & $< 25$ & $< 42$ & 74 & $60^{+10}_{-9}$ \\
$< 1.7$ & $< 24$ & $< 42$ & 66 & $55^{+9}_{-9}$ \\
$< 1.7$ & $< 24.5$ & $< 42.5$ & 71 & $60^{+10}_{-9}$ \\
$< 1.7$ & $< 24.5$ & $< 41.5$ & 67 & $58^{+10}_{-9}$ \\
$< 1.7$ & $< 24.5$ & $< 42$, upper limits $<42.5$ & 55 & $41^{+7}_{-7}$ \\
\enddata
\end{deluxetable}

We caution that the real uncertainties on the radio-quiet AGN
counts are larger than indicated owing to two effects, which go in opposite
directions: 1) $\sim 60\%$ of our putative radio-quiet AGN have elliptical
or undefined morphology, which means their optical fluxes have been
K-corrected assuming an elliptical spectrum. This could result in an
artificial lowering of $R$ if the AGN component is non negligible in the
optical band, translating in turn in an overestimation of their number (as
we selected sources with $R < 1.4$). Our selection criteria could also
include some radio galaxies, although their number is expected to be small;
2) the inclusion of X-ray detections only could underestimate the number of
radio-quiet AGN, as some sources with upper limits on their X-ray power
above $10^{42}$ ergs s$^{-1}$ could still have intrinsic powers above this
limit.

To assess the dependence of our results on our selection criteria we show
in Table \ref{perturb} how the fraction of sub-mJy SFG changes with
different parameters. Namely, we changed the values of $R$ (by $+0.6$,
$-0.3$), $P_{\rm r}$ (by $\pm0.5$ dex), and $L_{\rm x}$ (by $\pm 0.5$
dex). The fraction of sub-mJy SFG changes remarkably little, with a maximum
increase and decrease of $+3\%$ ($R = 2.3$) and $-6\%$ ($R = 1.4$)
respectively, well within the $1\sigma$ error for our standard set of
values. The biggest change happens when we include in our selection only
sources with upper limits on their X-ray powers $< 10^{42.5}$ ergs
s$^{-1}$, instead of all the X-ray non detections, to check for the effect
of our assumption that all sources with X-ray upper limits fall below the
AGN dividing line. The fraction of SFG decreases by almost one third to
$41\pm7\%$ (a $1.6\sigma$ difference), which shows that indeed our
assumption is a conservative one.

\subsection{The whole sample}\label{whole}

We now extend our analysis to the whole CDFS-VLA sample. As discussed in
\S~ \ref{CDFS-VLA} and \S~\ref{xrayinfo} (see also Figs. \ref{rloglr} and
\ref{rloglx}), in fact, the bulk of the sources without redshift are
unlikely to be SFG. Therefore, it would be unwise to simply extrapolate the
results of the redshift sub-sample to the whole sample. For the same
reason, any comparison with predicted number counts needs to be done with
the full sample.

We then apply our selection criteria described in \S~ \ref{redshift} to all
our sources, making various assumptions on the missing redshifts. Table
\ref{perturbz} shows that the fraction of sub-mJy SFG is only weakly
dependent on the assumed redshift for the sources without spectroscopic and
photometric redshifts. This is due to the fact that the increase in power
for larger redshifts is compensated by the decrease in $R$ due to the
larger K-correction, since we assumed elliptical morphology for the sources
without that information. For $z = \langle z \rangle$ the fraction of
sub-mJy SFG equals $57\pm8\%$ and in the following we make this assumption.

\begin{deluxetable}{cll}
\tabletypesize{\scriptsize}
\tablecaption{Fraction of sub-mJy star forming galaxies for different
  assumed redshifts. \label{perturbz}}
\tablewidth{0pt}
\tablehead{redshift & N &  sub-mJy fraction \\
     &                       &  \% \\}
\startdata
$\langle z \rangle$ & 88 & $57\pm8$ \\
1.0                             & 90 & $58\pm8$ \\
1.5                             & 89 & $58\pm8$ \\
2.0                             & 82 & $56\pm8$ \\
\enddata
\end{deluxetable}

\begin{deluxetable*}{lrccrcr}
\tabletypesize{\scriptsize}
\tablecaption{Euclidean normalized 1.4 GHz counts for the whole
  sample. \label{tab_counts}}
\tablewidth{0pt}
\tablehead{Flux range & Mean Flux Density &   & & Counts &  & \\
                 & & ##Total & SF & fraction & AGN & fraction \\
###$\mu$Jy & $\mu$Jy & sr$^{-1}$ Jy$^{1.5}$ & sr$^{-1}$ Jy$^{1.5}$ & \% &
  sr$^{-1}$ Jy$^{1.5}$ & \% \\
}
\startdata
$##43 - 75$ & #63 &$#2.53^{+0.51}_{-0.42}$ & $1.61^{+0.41}_{-0.33}$ &
$64^{+19}_{-18}$ & $0.92^{+0.37}_{-0.27}$ & $36^{+16}_{-13}$ \\
$##75 - 120$ & #97 &$#2.62^{+ 0.45}_{-0.38}$ & $1.69^{+0.38}_{-0.31}$ &
$65^{+17}_{-16}$ & $0.87^{+0.28}_{-0.22}$ & $33^{+12}_{-10}$ \\
$#120 - 200$ & 152 & $#2.87^{+0.53}_{-0.45}$ & $1.03^{+0.36}_{-0.27}$ &
$36^{+14}_{-12}$ & $1.68^{+0.42}_{-0.34}$ & $59^{+17}_{-16}$ \\
$#200 - 500$ & 306 & $#4.02^{+0.74}_{-0.63}$ & $1.81^{+0.53}_{-0.42}$ &
$45^{+15}_{-13}$ & $2.11^{+0.57}_{-0.46}$ & $52^{+16}_{-15}$ \\
$#500 - 2000$ & 1032 & $#6.71^{+1.86}_{-1.49}$ & $1.34^{+1.06}_{-0.64}$ &
$20^{+16}_{-11}$ & $5.37^{+1.70}_{-1.33}$ & $80^{+31}_{-30}$ \\
$2000 - 100000$ & 17262 &$42.4^{+12.5}_{-9.9}$ & 0.0 & 0## &
$42.4^{+12.5}_{-9.9}$ & $100^{+38}_{-38}$\\
\enddata
\end{deluxetable*}

\begin{figure}
\centerline{\includegraphics[width=9.5cm]{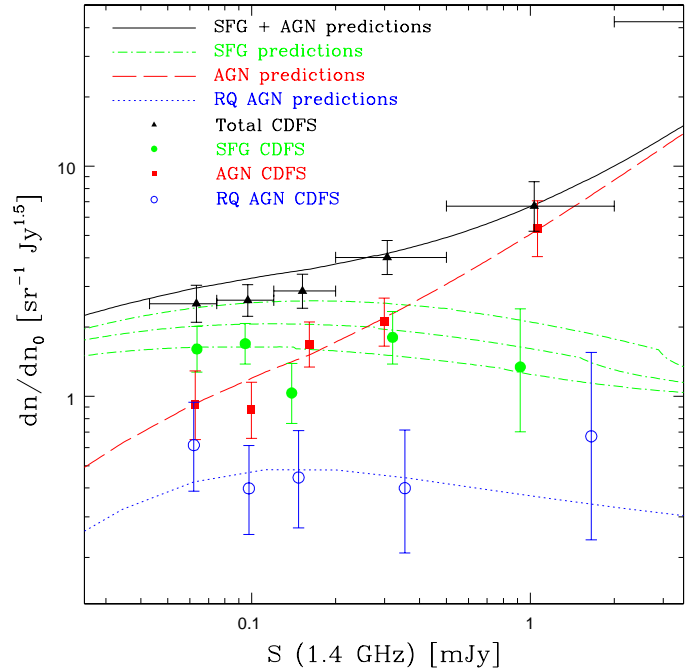}}
\caption{The Euclidean normalized 1.4 GHz CDFS source counts: total counts
  (black triangles), SFG (filled green circles), all AGN (red squares), and
  radio-quiet AGN (open blue circles). Error bars correspond to $1\sigma$
  errors \citep{geh86}. Model calculations refer to SFG (green dot-dashed lines),
  displayed with a $1\sigma$ range on the evolutionary parameters, all AGN
  (red dashed line), radio-quiet AGN (blue dotted line), and the sum of the first two
  (black solid line). See text for more details.} \label{counts}
\end{figure}

Table \ref{tab_counts} and Fig. \ref{counts} present the Euclidean
normalized number counts for the whole sample and also for SFG and AGN,
derived as described in Paper I.  The trends are very similar to those of
the redshift sub-sample. AGN make up $41\pm6\%$ of sub-mJy sources and their
counts are seen to drop at lower flux densities, going from 100\% of the total 
at $\sim 10$ mJy down to $36\%$ at the survey limit. SFG, on the other
hand, which represent $57\pm8\%$ of the sample, are missing at high flux
densities but become the dominant population below $\approx 0.1$ mJy,
reaching $64\%$ at the survey limit. Radio-quiet AGN represent
$19^{+5}_{-4}\%$ (or $46\%$ of all AGN) of sub-mJy sources but their
fraction appears to increase at lower flux densities, where they make up
$66\%$ of all AGN and $\approx 24\%$ of all sources at the survey
limit, up from $\approx 10\%$ at $\sim 1$ mJy.

\subsection{Expected number of radio sources}\label{expected}

In Fig. \ref{counts} we overlay some model calculations for the AGN and SFG
contribution to the radio source counts. The AGN model includes FR I and FR
II radio galaxies and BL Lacertae objects, the only populations
significantly present in the counts below a few mJy at 1.4 GHz (Padovani,
in preparation). In fact, according to unified schemes, radio quasars,
being FR IIs with their powers boosted by relativistic beaming, are
expected to disappear from the radio population below $S_{\rm r} \la 1$ mJy
\citep{pad92}.

Radio galaxy counts were calculated starting from the 2.7 GHz luminosity
functions of 2 Jy FR I and II radio galaxies derived by \cite{up95},
converted to 1.4 GHz assuming $\alpha_{\rm r} = 0.7$. Based on their
results, no evolution was assumed for FR Is, while their best fit
luminosity evolution ($\tau = 0.26$ for an exponential evolution $P(z) =
P(0) \exp[T(z)/\tau]$, where $T(z)$ is the look-back time and $\tau$ is the
evolutionary time scale in units of the Hubble time) was used for FR
IIs. BL Lac number counts were estimated from the beaming model of
\cite{up95} (converted from 5 GHz assuming $\alpha_{\rm r} = -0.1$), which
fits very well not only the original 1 Jy data on which it was based but
also the deeper ($\sim 50$ mJy) DXRBS sample \citep{pad07a}.

The radio AGN counts are dominated by FR I radio galaxies at sub-mJy
levels. In fact, given their higher radio power and stronger evolution, FR
IIs reach only $S_{\rm r} \approx 0.5$ mJy. FR Is appear to make up $\sim
40\%$ of the radio galaxies with $S_{\rm r} > $1 mJy but $\sim 90\%$ of the
whole sample, and to reach flux densities as low as $\sim 0.1~\mu$Jy. BL
Lacs can also reach quite faint flux densities ($\sim 10~\mu$Jy) but make
up only a minority of the AGN, $\approx 10 - 15\%$ in the range $0.1 - 1$
mJy.

The surface density for SFG has been derived following \cite{hop04}.
Namely, the local luminosity function from \cite{sad02} was adopted, with
luminosity evolution $\propto (1+z)^{2.7}$ to $z = 2$ (and constant
thereafter) and density evolution $\propto (1+z)^{0.15}$. To have a feeling
for the likely uncertainties, we have also evaluated the number counts by
varying the evolutionary parameters by $1 \sigma$.

Our AGN counts include a radio-quiet component, as radio-quiet AGN are not
radio-silent and are routinely detected at faint flux densities in the
radio band.  To estimate their expected number counts one would need a
sizable sample of such {\it radio-selected} sources from which to derive
the radio luminosity function and evolution, to then predict their number
densities at lower flux densities, as we have done for radio- and star
forming galaxies. Unfortunately, such a sample is not available.
Radio-quiet AGN, however, are very strong X-ray emitters and indeed
dominate the X-ray sky. We therefore took the $5 - 10$ keV AGN number
counts given by \cite{com08} \citep[based on][]{gil07}, which fit the
observed X-ray counts down to $\sim 10^{-15}$ ergs cm$^{-2}$ s$^{-1}$ and
make predictions down to $\sim 10^{-17}$ ergs cm$^{-2}$ s$^{-1}$, to obtain
an estimate of the radio-quiet AGN counts in the radio band. We first
multiplied the X-ray counts by 0.9, to take into account the fact that
radio-loud AGN make up $\approx 10\%$ of all AGN, and then converted them
to Euclidean normalized counts. To derive radio number counts one would
then need to know the {\it intrinsic} radio-to-Xray flux ratio
distribution, or at least the mean ratio for radio-quiet AGN. This is at
present unknown. However, one can obtain robust bounds on this ratio by
looking at X-ray selected samples and our own sources.

\cite{pol07} present the SEDs of a hard X-ray selected sample of AGN with
$f_{\rm 2 - 10 keV} > 10^{-14}$ ergs cm$^{-2}$ s$^{-1}$. These include
radio flux densities, mostly upper limits. We used survival analysis to
derive a k-corrected mean value $\langle S_{1.4GHz} /f_{\rm 2 - 10 keV}
\rangle = 14.57\pm0.20$ (where the radio flux density is in $\mu$Jy and the
X-ray flux is in c.g.s. units). Radio-loud AGN, which we define as sources
with $L_{\rm r} > 10^{24}$ W Hz$^{-1}$ and make up $\sim 13\%$ of the
sample, were excluded. The radio-quiet AGN in our complete sample, defined
as done in \S~\ref{redshift}, have $\langle S_{1.4GHz} /f_{\rm 2 - 10 keV}
\rangle = 16.72\pm0.16$. Both of these values are heavily influenced by
selection effects. X-ray selection will tend to favor sources with
relatively small radio-to-X-ray flux ratios, while the opposite will be
true for radio selection. The truth, as always, will be somewhere in the
middle and actually quite far from the two extremes. We then take as our
best guess for the radio-to-X-ray flux ratio for radio-quiet AGN the mean
of the two values, that is $\langle S_{1.4GHz} /f_{\rm 2 - 10 keV} \rangle
= 15.6$, being fully aware that this will only give an order of magnitude
estimate. We note that this value is the same as that obtained from the
sample of \cite{bri00} by converting their 2 keV powers to the $2 - 10$ keV
band assuming $\alpha_{\rm x} = 0.8$. Establishing why this is so, given
the fact that the \cite{bri00} sample was derived from a cross-correlation
between ROSAT All-Sky Survey and 1.4 GHz FIRST data, is beyond the scope of
this paper. We just remark that this validates the usage of this value as
typical in the literature \citep[e.g.,][]{sim06,pol07}.  Radio number
counts for radio-quiet AGN were then obtained by using the hard X-ray
counts and the value derived above converted to $5 - 10$ keV by assuming
$\alpha_{\rm x} = 0.8$ (i.e., $\langle S_{1.4GHz} /f_{\rm 5 - 10 keV}
\rangle = 16.0$).

The agreement between observed and expected number counts for the various
sub-classes and for the whole sample in Fig. \ref{counts} is very good,
especially given the many orders of magnitude difference between our flux
density limit and those of the samples from which the luminosity functions
were derived. Even the radio-quiet AGN counts show a very good agreement
with our observational estimates. We point out that the inclusion of a
distribution in radio-to-X-ray flux ratio would "stretch" those counts
conserving the total number, with the result that the drop at lower flux
densities would be somewhat mitigated.

\subsection{Heavily absorbed AGN: the meaning of $R$}

\begin{figure}
\centerline{\includegraphics[width=9.5cm]{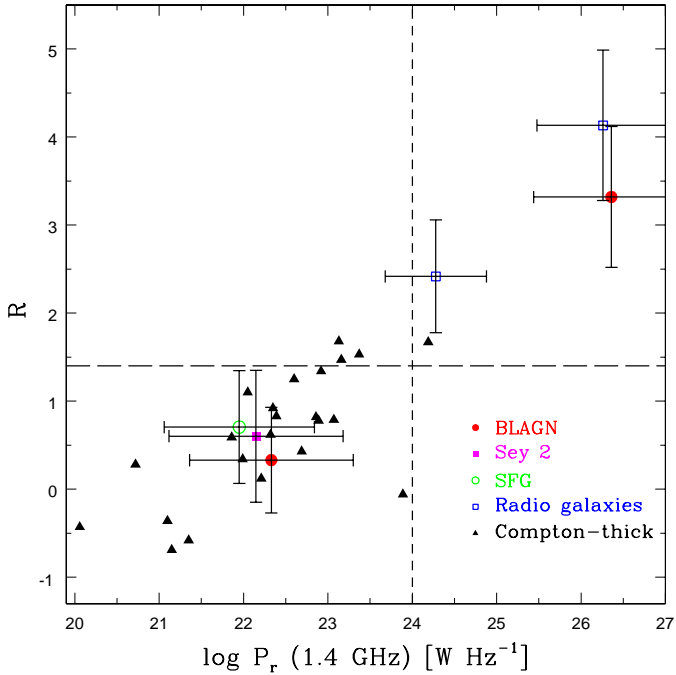}}
\caption{The radio-to-optical flux density ratio $R = \log (S_{\rm
    1.4GHz}/S_{\rm V})$ versus the 1.4 GHz radio power for various classes
  of sources, from the AGN catalogue of \cite{pad97}: Broad Line AGN
  (BLAGN), radio-quiet (left-most point) and radio-loud (right-most point),
  filled red circles; Seyfert 2s, filled magenta square; SFG, open green
  circle; and radio galaxies with $\log P_{\rm r} \le 25$ (left-most point)
  and $\log P_{\rm r} > 25$ (right-most point), open blue
  squares. Individual data for $\sim 55\%$ of the Compton-thick sources in
  the compilation of \cite{com04} are also shown (black triangles). Error
  bars represent the $1\sigma$ dispersion. The horizontal (long-dashed)
  line at $R = 1.4$ and the vertical (short-dashed) line at $\log P_{\rm r}
  = 24$ indicate the approximate maximum values for local SFG and dividing
  lines between radio-loud and radio-quiet AGN. See text for more
  details.}  \label{rloglr_mean}
\end{figure}

Radio selection gives us the chance of identifing obscured AGN which might
escape X-ray detection \citep[see, e.g.,][]{don05}. One might think that,
since their optical flux is heavily absorbed, such sources will have quite
large radio-to-optical flux density ratios. Indeed, it has been suggested
that the occurrence of such large ratios in micro-Jy sources is indicative
of a high degree of extinction \citep[e.g.,][]{nor05}. Radio-quiet,
obscured AGN (the so-called "type 2" AGN), would then appear to have $R$
values very large for their radio power and should stick out in a $R$ -
radio power plot. To test this hypothesis we plot in Fig. \ref{rloglr_mean}
mean $R$ and $\log P_{\rm r}$ values for broad-lined AGN, both radio-quiet
and radio-loud, Seyfert 2s, and star forming and radio galaxies. The data
come from the AGN catalogue of \cite{pad97}. Given the redshift
distribution of our sample (\S~\ref{CDFS-VLA}) and to compare sources within
a similar redshift range, we have restricted our selection to $z \le
1$. Individual data from the same catalogue\footnote{For consistency with
our sources the optical magnitudes used in the evaluation of the
radio-to-optical flux density ratio refer, as much as possible, to the
total, and not the nuclear, magnitude.} for $\sim 55\%$ of the Compton-thick
sources from the compilation of \cite{com04} are also shown. In such
sources, defined as having hydrogen column values $N_{\rm H} \ga1.5 \times
10^{24}$ cm$^{-2}$, nuclear radiation is only visible above $\sim 10$ keV
for $N_{\rm H} \la 10^{25}$ cm$^{-2}$, the so-called "mildly Compton thick"
sources, while for larger column densities the entire high-energy X-ray
spectrum is absorbed. Compton-thick sources are expected to be heavily
absorbed in the optical band as well, and indeed $\sim 80\%$ of them are
classified as Seyfert 2s/QSO 2s or radio galaxies.

An important result is apparent from Fig. \ref{rloglr_mean}: X-ray and
optically absorbed sources fall on the same relation as optically
unabsorbed ones. In other words, a large value of the radio-to-optical flux
density ratio does not indicate a high degree of extinction but, rather, a
large radio power. A similar story is told by the Compton-thick sources,
which fall on the same relation as the other sources (see also Fig. 
\ref{rloglr}).  This can be explained by the fact that there is a limit to
how faint the optical flux will be, since once the central AGN is
completely obscured the host galaxy will contribute all of the
flux. Radio-quiet, broad-lined AGN have $R$ values only slightly smaller
than Seyfert 2s because, given also the redshift cut-off, they are mostly
nearby and therefore with a strong galaxy component. Radio-quiet quasars at
all redshifts, which have a stronger AGN-related optical component, have
$\langle R \rangle \sim 0.54$ and $\log P_{\rm r} \sim 24.1$, that is below
the locus of the other sources.

We notice that two sources in Fig. \ref{rloglr} (top-left quadrant) appear
to be off the main relation by having $R$ values larger than expected for
their radio powers. These are at low redshift ($z \sim 0.06$ and 0.18). One
of them has been studied by \cite{toz06} (ID 97 in their Table 1), who
derived $\alpha_{\rm x} = 0.30^{+0.11}_{-0.10}$ and $N_{\rm H} \sim 0$.
The other one is a much weaker X-ray source with an X-ray hardness ratio,
defined as $HR = (H - S)/(H+S)$, where $H$ and $S$ are the net counts in
the hard and soft band respectively, $\sim 0.41$. This implies $N_{\rm H}
\ga 10^{22}$ cm$^{-2}$. In both cases, then, there is no strong indications
of absorption. Both sources have very faint optical magnitudes for their
redshifts, clearly off the locus of the majority of our sources, which
explains their high $R$. This suggests that they might be somewhat
peculiar.

Based on all of the above, it appears that radio and optical data cannot
identify heavily absorbed sources, which might have escaped X-ray
detection. This can be investigated further using Spitzer data.

\section{Discussion}

By using a variety of multi-wavelength information for 256 radio sources in
the CDFS area, we have shed light on the long standing problem of the
nature of sub-mJy radio sources. We find that SFG make up $\la 60\%$ of
such sources. We believe our results are quite robust, since whenever we
made some assumptions, we chose those which would {\it maximize} the number
of SFG. AGN make up the remaining fraction ($\ga 40\%$) of radio sources
with 1.4 GHz flux densities $< 1$ mJy.  The observational evidence, backed
up by model calculations and an extrapolation from the hard X-ray band,
strongly supports the suggestion that AGN are roughly equally split into
radio-loud, mostly low-power radio galaxies, and radio-quiet sources.

The population mix appears to be flux density dependent, with SFG becoming
increasingly relevant at lower flux densities until they become the
dominant component below $\approx 0.1$ mJy, and AGN exhibiting the opposite
trend. This seems to be driven by the fall of radio-loud sources, as
radio-quiet ones display relatively flat counts (within the rather large
errors). SFG do appear to be responsible for most of the well-known up-turn
in the counts, with radio-quiet AGN providing only a relatively small
($\approx 25\%$) contribution.

The availability of optical morphology and X-ray information (including
upper limits) for $\sim 70\%$ and all of the sources in the complete sample
respectively have been fundamental to classify our sources. X-ray data, in
particular, have been vital to eliminate AGN contaminants from spirals and
irregulars and to classify sources without morphological information. We
have obviated the presence of many X-ray upper limits by using them
conservatively and thereby providing robust upper limits to the fraction of
SFG. Deeper X-ray data should allow us to better define the fraction of
SFG.

As mentioned in the Introduction, our results differ somewhat with the many
papers, which over the years have suggested a large dominance of SFG at
sub-mJy levels. Many of these results were based on only a small fraction
of spectroscopic identifications of the optically brightest
objects. \cite{gru99} and \cite{cil03} have pointed out that, when
identifying spectroscopically fainter objects, early-type galaxies
constitute the majority. This is easily explained: if we cut our sample at
$V < 22$, for example, we would be sampling a region of parameter space
typical of SFG, with $R \sim 0.5$, $z \sim 0.5$, $\log P_{\rm r} \sim 23$,
and $\log L_{\rm x} \sim 41$ (see Figs. \ref{vfrcdfs}, \ref{rloglr}, and
\ref{rloglx}). We would then wrongly infer that the sub-mJy population does
not include the large number of AGN, which become accessible only when
probing fainter magnitudes.

Our results are in broad agreement with a number of recent papers. For
example, \cite{cil03} have obtained a deep 5 GHz VLA image of the Lockman
hole. The analysis of their 63 radio sources, which reach $\sim 50~\mu$Jy,
using a color-color diagram and a radio flux density - optical magnitude
plot, suggests that AGN are still the dominant population at their flux
density levels.

\cite{smo08} have applied a new method to disentangle star forming from AGN
galaxies using $\sim 2,400$ 1.4 GHz radio sources in the VLA-COSMOS
survey. This relies on a correlation between optical rest-frame colors of
emission-line galaxies and their position in a spectroscopic diagnostic
diagram based on emission line flux ratios. They deduce that the radio
population in the flux density range $50~ \mu$Jy $- 0.7$ mJy is a mixture
of $30 - 40\%$ SFG, $50 - 60\%$ AGN galaxies, and $\sim 10\%$
QSO, where the latter are defined as optical point sources. Their method
works only up to $z \sim 1.3$ and therefore beyond that value their results
are less secure.

\cite{sey08} have used a variety of (mostly) radio-based diagnostics,
including radio morphology, spectral indices, radio to near-IR and mid-IR
flux density ratios, to discriminate on a statistical basis AGN from SFG in
a deep VLA/MERLIN survey of the $13^H$ {\it XMM-Newton/Chandra} Deep Field,
which includes 449 sources. They find a dominance of SFG at the faintest
flux densities ($\sim 50~\mu$Jy at 1.4 GHz), with AGN still making up around
one quarter of the counts. They also find that the relative contribution of
the two classes is flux density dependent, with SFG becoming the more
numerous population below $\sim 0.1$ mJy.

It is important to note that our selection criteria, by being not only
radio-based, provide information on the {\it overall} nature of our
sources, unlike, for example, those of \cite{sey08} \citep[see
  also][]{ric07}, which separate radio sources into those whose {\it radio
  emission} is AGN and star formation dominated.
 
Regarding radio-quiet AGN, \cite{jar04} modeled the radio counts with a
population of radio-quiet AGN becoming the dominant component below 1
mJy. We do not confirm this prediction. \cite{sim06}, on the other hand,
suggested that $\ga 20\%$ of radio sources in the $0.1 - 0.3$ mJy range
were radio-quiet AGN and that these sources could be more important than
SFG for the up-turn in the counts. While our results are consistent with
the first point, they are not with the second one.

We notice that radio-quiet AGN counts appear to be similar in shape to
those of SFG, scaled down by a factor $\approx 3 - 4$ (see
Fig. \ref{counts}), and quite different from those of the whole AGN
population, which includes basically only radio-loud sources above 1
mJy. This argues against radio-quiet AGN being simply the faint end of a
continuum of radio powers and should provide relevant information for the
solution of the puzzle of the origin of their radio emission.

\section{Summary and Conclusions}

We have used a deep radio sample, which includes 256 objects down to a 1.4
GHz flux density of $42~\mu$Jy selected in the Chandra Deep Field South
area, to study the nature of sub-mJy sources. Our unique set of ancillary
data, which includes reliable optical/near-IR identifications, optical
morphological classification, redshift information, and X-ray detections or
upper limits for a large fraction of our sources, has allowed us to shed
new light on this long standing astrophysical problem. By analyzing the
ratio of radio to optical luminosity and the radio and X-ray powers of the
sources with morphological and redshift information, we have selected
candidate star-forming galaxies and AGN from a complete sample of 194
objects. As the first two parameters by themselves are not very good
discriminants between star-forming galaxies and AGN, optical morphology and
especially X-ray data turned out to be vital in establishing the nature of
faint radio sources. We have also complemented our data analysis with model
calculations based on brighter samples and counts in other bands. Our main
results can be summarized as follows:

\begin{enumerate}
\item The well-known flattening of the radio number counts below $\approx
  1$ mJy is mostly due to star-forming galaxies, which are missing above
  $\sim 2$ mJy but become the dominant population below $\approx 0.1$ mJy.
\item AGN exhibit the opposite behavior, as their counts drop at lower flux
  densities, going from 100\% of the total at $\sim 10$ mJy down to $<
  50\%$ at the survey limit. This is driven by the fall of radio-loud
  sources, as radio-quiet objects, which make up 
  $\sim 20\%$ of sub-mJy sources, display relatively flat counts.
\item Radio-quiet AGN make up about half of all AGN. Their counts appear to be a scaled down version, by a
  factor $\approx 3 - 4$, of those of star-forming galaxies, and are very
  different from those of the radio-loud AGN population. This should
  provide a clue to the origin of their radio emission.
\item Star-forming galaxies make up $\la 60\%$ of sub-mJy sources down to
  the flux limit of this survey. This has to be regarded as a robust upper
  limit, as whenever we had to make some assumptions, we chose to maximize
  their numbers. This result is at variance with the many papers, which
  over the years have suggested a much larger dominance of these
  sources. On the other hand, our results are in broad agreement with a
  number of recent papers, which found a significant AGN component down to
  $\approx 50~\mu$Jy.
\item The results of our model calculations agree quite well with the
  observed number counts and provide supporting evidence for the scenario
  described above. Moreover, they imply that sub-mJy radio-loud AGN are
  dominated by low-power, Fanaroff-Riley type I radio galaxies, as their
  high-power counterparts and radio-loud quasars are expected to disappear
  below $\sim 0.5 - 1$ mJy.
\item We find a correlation between X-ray and radio power for star-forming
  galaxies, in agreement with most previous studies.
\item Despite some previous claims, we find that the ratio of radio to
  optical luminosity depends more on radio luminosity, rather than being
  due to optical absorption, and is therefore not related to a high degree
  of extinction.
\end{enumerate}

Considering the apparent emerging population of low luminosity AGN at
microjansky levels, care is needed when interpreting radio source counts in
terms of the evolution of the star formation rate in the Universe.

We plan to expand on this work by using our deeper ($7~\mu$Jy per beam over
the whole Extended CDFS region) radio observations \citep{mil08} and the
recently released 2 Msec Chandra data \citep{luo08}.

\begin{acknowledgements}
NM acknowledges the support of a Jansky Fellowship. PT acknowledges the
financial support of contract ASI--INAF I/023/05/0. We thank Andrea
Comastri and Mari Polletta for providing data from their papers in
electronic format.  We acknowledge the ESO/GOODS project for the ISAAC and
FORS2 data obtained using the Very Large Telescope at the ESO Paranal
Observatory under Program ID(s): LP168.A-0485, 170.A-0788, 074.A-0709, and
275.A-5060.  The VLA is a facility of the National Radio Astronomy
Observatory which is operated by Associated Universities, Inc., under a
cooperative agreement with the National Science Foundation. This research
has made use of the NASA/IPAC Extragalactic Database (NED) which is
operated by the Jet Propulsion Laboratory, California Institute of
Technology, under contract with the National Aeronautics and Space
Administration.

{\it Facilities:} \facility{VLA}, \facility{ESO:3.6m (EFOSC2)},
\facility{VLT:Kueyen (FORS1)}, \facility{VLT:Antu (FORS2)},
\facility{VLT:Antu (ISAAC)}, \facility{Max Plank:2.2m (WFI)},
\facility{CXO}, \facility{XMM}

\end{acknowledgements}

\end{document}